\documentclass[aip,amsmath,amssymb,reprint,floatfix]{revtex4-1}
\usepackage{graphicx}
\usepackage{bm}
\usepackage[hidelinks]{hyperref}
\usepackage[T1]{fontenc}
\usepackage[utf8]{inputenc}
\usepackage{mathptmx}
\usepackage{braket}

\begin{document}
\title{Axis-selective AC magnetic-field imaging in the tens-of-megahertz regime using RF-dressed continuous-wave scanning NV magnetometry}

\author{S. Komuro}
\affiliation{School of Fundamental Science and Technology, Keio University, Yokohama, Kanagawa \mbox{223-8522, Japan}}
\affiliation{Center for Spintronics Research Network, Keio University, Yokohama, Kanagawa \mbox{223-8522, Japan}}
\author{R. Okaniwa}
\affiliation{School of Fundamental Science and Technology, Keio University, Yokohama, Kanagawa \mbox{223-8522, Japan}}
\affiliation{Center for Spintronics Research Network, Keio University, Yokohama, Kanagawa \mbox{223-8522, Japan}}
\author{K. Okura}
\affiliation{School of Fundamental Science and Technology, Keio University, Yokohama, Kanagawa \mbox{223-8522, Japan}}
\affiliation{Center for Spintronics Research Network, Keio University, Yokohama, Kanagawa \mbox{223-8522, Japan}}
\author{K. Hayashi}
\affiliation{School of Materials Science, Japan Advanced Institute of Science and Technology, Nomi \mbox{923-1292, Japan}}
\author{T. An}
\affiliation{School of Materials Science, Japan Advanced Institute of Science and Technology, Nomi \mbox{923-1292, Japan}}
\author{J. Ishi-Hayase}
\email{hayase@appi.keio.ac.jp}
\affiliation{School of Fundamental Science and Technology, Keio University, Yokohama, Kanagawa \mbox{223-8522, Japan}}
\affiliation{Center for Spintronics Research Network, Keio University, Yokohama, Kanagawa \mbox{223-8522, Japan}}

\begin{abstract}
Spatially resolved magnetic-field imaging reveals material properties and local currents in microelectronic devices. Conventional pulsed scanning nitrogen-vacancy (NV) magnetometry (SNVM) faces increasingly stringent microwave (MW) pulse-control requirements for detecting radio-frequency (RF) magnetic fields at frequencies ranging from tens to hundreds of megahertz. To address this challenge, we implement an RF-dressed continuous-wave (CW) sensing scheme for SNVM and demonstrate AC magnetic-field imaging at 33.5~MHz. The method exploits electron-spin double resonance in the NV spin-1 three-level system: the target RF magnetic field coherently couples two spin eigenstates, forming RF-dressed states whose splittings are detected in continuous-wave electron-spin double resonance (CW-ESDR) spectra under continuous laser, MW, and RF excitation. This approach avoids short, high-power MW pulses and synchronization between the RF field and a pulsed sensing sequence. Using an NV-ensemble diamond probe integrated with an atomic force microscope (AFM), we image the Oersted field of an RF-driven L-shaped microcircuit. Successively selecting three crystallographic NV axes by static bias-field control yields maps of RF magnetic-field projection magnitudes, with distinct spatial distributions qualitatively reproduced by Biot--Savart calculations. Separately, transverse bias-field control tunes the RF-coupled spin-state transition frequency from 14.4 to 89.3~MHz, as characterized by spectroscopy and Hamiltonian analysis, with RF-dressed resonances directly observed in CW-ESDR spectra at 89.3~MHz. Together with previous RF magnetic-field detection up to 632~MHz using the same RF-dressed CW sensing principle in a confocal NV setup, these results support extending AC magnetic-field imaging with SNVM toward the hundreds-of-megahertz regime and, with additional phase information, vector magnetic-field reconstruction.
\end{abstract}
\maketitle

\section{Introduction}
\label{sec:Introduction}
Spatially resolved magnetic-field imaging is a powerful tool for investigating the physical properties of materials and diagnosing local current distributions in increasingly miniaturized electronic devices, particularly when conventional terminal measurements provide only spatially averaged information and cannot resolve local current pathways. Established magnetic imaging techniques include scanning superconducting quantum interference device (SQUID) microscopy, scanning Hall-probe microscopy, and magnetic force microscopy. \cite{Kirtley2010,Gerken2020,Winkler2023} These techniques, however, involve trade-offs among magnetic sensitivity, spatial resolution, operating conditions, and measurement invasiveness. \cite{Christensen2024}

Diamond nitrogen-vacancy (NV) centers have attracted considerable attention for room-temperature magnetic-field imaging owing to their high magnetic sensitivity and potential for high spatial resolution. \cite{Degen2017} The magnetic-field response can be detected through optically detected magnetic resonance (ODMR) of the NV electron spins. Among NV-based imaging approaches, scanning NV magnetometry (SNVM) is particularly powerful for achieving nanoscale spatial resolution by combining an NV-containing diamond probe with atomic force microscopy (AFM). \cite{Maletinsky2012} SNVM has been successfully used to map current distributions in nanoscale conductors and graphene devices, \cite{Chang2017,Palm2024} and more recently to investigate current flow through conducting ferroelectric domain walls and graphene--metal interfaces. \cite{McCluskey2025,Ding2026} Reduced sensor--sample separation and multiplexed sensing with dual-NV probes have further extended the capabilities of SNVM. \cite{Xu2025,Huxter2025}

Pulsed SNVM has imaged magnetic fields generated by AC currents at 1.33~MHz, \cite{Palm2022} whereas Rabi-based scanning measurements have mapped microwave (MW) magnetic fields in the gigahertz regime by resonantly driving NV electron-spin transitions. \cite{Appel2015,Rable2025,LiResonator2026} However, experimental AC magnetic-field imaging in the intermediate frequency range between 10~MHz and the gigahertz regime had not previously been demonstrated with SNVM. Outside scanning-probe configurations, pulsed wide-field NV magnetometry has demonstrated radio-frequency (RF) magnetic-field imaging at 19.23~MHz using multipulse dynamical decoupling (DD) and up to 70~MHz using quantum frequency mixing. \cite{Nomura2023,Karlson2024} Both demonstrations, however, relied on pulsed spin-interrogation protocols.

Although pulsed DD magnetometry using Carr--Purcell--Meiboom--Gill (CPMG) and XY-$n$ sequences is widely used for frequency-selective detection of high-frequency AC magnetic fields, its implementation becomes increasingly challenging at higher detection frequencies. These pulsed protocols exploit coherent phase accumulation, with the interpulse spacing set by the period of the target AC field. \cite{Pham2012,Degen2017} At detection frequencies of several tens of megahertz and above, the required short interpulse intervals demand correspondingly short MW $\pi$ pulses and high Rabi frequencies, making precise pulse timing and spin control increasingly demanding. \cite{Ishikawa2018,Ghimire2023,Jung2025} Furthermore, phase-sensitive implementations require synchronization between the target AC magnetic field and the MW pulse sequence.

These experimental requirements motivate a complementary continuous-wave (CW) approach based on electron-spin double resonance (ESDR). In the transverse-bias-field configuration, a static bias magnetic field is applied nearly perpendicular to the selected NV axis, mixing the $m_s=\pm1$ sublevels into two upper spin eigenstates. The target RF magnetic field resonantly couples these eigenstates and forms RF-dressed states whose amplitude-dependent splitting can be detected in CW-ESDR spectra under continuous laser, MW, and RF excitation. This enables MHz-range AC magnetic-field sensing without the short, high-power MW $\pi$ pulses or RF-to-pulse-sequence synchronization required in conventional pulsed high-frequency AC magnetometry. Saijo \textit{et al.} demonstrated AC magnetic-field sensing using this approach. \cite{Saijo2018}

The transverse bias field also enables spectral selection of a specific crystallographic NV orientation from the four possible NV axes, as demonstrated by Yamaguchi \textit{et al.} \cite{Yamaguchi2019} In addition, this configuration can reduce the first-order sensitivity of the relevant transition to fluctuations in the magnetic-field component parallel to the selected NV axis, while RF dressing suppresses inhomogeneous broadening arising from strain and electric-field variations. RF-dressed linewidth narrowing has been demonstrated in bulk diamond \cite{Tabuchi2023} and in nanodiamonds. \cite{Beaver2024} The ability to suppress such resonance broadening is particularly beneficial for nanodiamonds and other small diamond probes, which can exhibit broader resonances owing to strain and surface-related inhomogeneities. Furthermore, the RF-coupled transition frequency can be tuned by controlling the transverse bias field. Following frequency-tunable ESDR using additional RF dressing, \cite{Okaniwa2024} AC magnetic-field detection from 14.4 to 632~MHz was demonstrated in a confocal NV setup using transverse-bias-field control. \cite{Okaniwa2026arxiv}

Here, we demonstrate axis-selective AC magnetic-field imaging in the tens-of-megahertz regime using RF-dressed CW SNVM, a frequency regime that had not previously been accessed by AC magnetic-field imaging with SNVM. Using an NV-ensemble diamond probe integrated with an AFM, we image the AC Oersted field generated by an L-shaped microcircuit driven at 33.5~MHz. By successively controlling the direction of the static bias field, three crystallographic NV orientations are selected without mechanically rotating either the probe or the sample circuit, yielding maps of the RF magnetic-field projection magnitude along each selected NV axis. The observed axis-dependent spatial distributions are qualitatively reproduced by Biot--Savart calculations. Separately, transverse-bias-field control tunes the RF-coupled spin-state transition frequency from 14.4 to 89.3~MHz, with RF-dressed resonances directly observed in CW-ESDR spectra at 89.3~MHz. These results extend AC magnetic-field imaging with SNVM into the tens-of-megahertz regime and provide a basis for future high-frequency, multi-axis, and vector RF magnetic-field imaging.

\section{Principle of RF-dressed CW magnetometry}
\label{sec:principle}

\begin{figure}[t]
\centering
\includegraphics[width=\columnwidth]{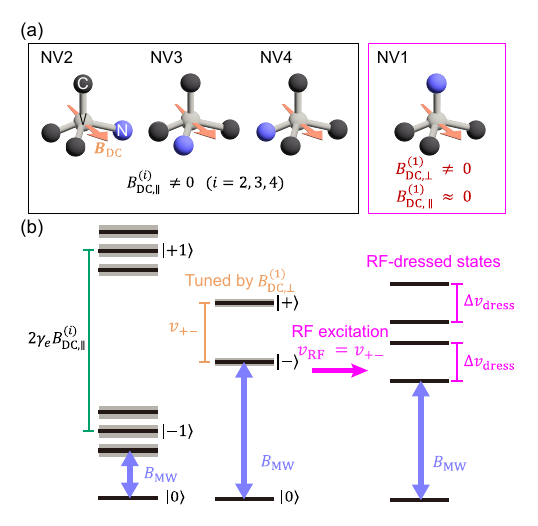}
\caption{Principle of frequency-tunable, axis-selective RF-dressed CW NV
magnetometry. (a) Schematic of the four crystallographic NV orientations and
the static bias magnetic field. In the illustrated configuration, the bias
field is nearly perpendicular to NV1, whereas the other orientations have
finite longitudinal field components. The parallel and transverse components
are defined relative to each NV axis.
(b) Schematic of the NV ground-state energy levels for orientations with
predominantly longitudinal and transverse bias-field components.
In the selected transverse-field configuration, the two upper spin
eigenstates $\ket{+}$ and $\ket{-}$ are resonantly coupled
by the RF magnetic field, producing RF-dressed splittings in the CW-ESDR
spectrum. The transition frequency between these states can be tuned by
adjusting the transverse bias-field magnitude.}
\label{fig:principle}
\end{figure}

The detection principle is illustrated in Fig.~\ref{fig:principle}.
The negatively charged NV center in diamond has four possible crystallographic orientations.
By adjusting the direction of the static bias magnetic field
$\boldsymbol B_{\mathrm{DC}}$, one NV orientation can be placed in a configuration in which
 the field is nearly perpendicular to its symmetry axis, while the remaining orientations
experience appreciable longitudinal field components.
This configuration enables spectral selection of the NV orientation used for RF magnetic-field sensing.
Figure~\ref{fig:principle}(a) depicts this arrangement with NV1 as the selected orientation;
NV1, NV2, and NV3 are selected sequentially in the experiments discussed below.
The energy-level structures corresponding to transverse- and longitudinal-field-dominated
configurations are illustrated in Fig.~\ref{fig:principle}(b).

We use a laboratory coordinate system $(x,y,z)$, where $x$ and $y$ lie in the
sample plane and $z$ is normal to it.
The four crystallographic NV orientations are represented in this frame by unit
vectors $\boldsymbol n_{\mathrm{NV},i}=(n_{x,i},n_{y,i},n_{z,i})^{\mathsf{T}}$,
$i=1,2,3,4$.
For the $i$th orientation, the local coordinate system
$(x'_i,y'_i,z'_i)$ has $z'_i$ parallel to $\boldsymbol n_{\mathrm{NV},i}$;
$x'_i$ and $y'_i$ lie in the plane perpendicular to that NV axis.
The signed longitudinal component and the transverse-field magnitude
relative to the $i$th axis are defined as

\begin{align}
B_{\mathrm{DC},\parallel}^{(i)}
 &=\boldsymbol B_{\mathrm{DC}}\cdot\boldsymbol n_{\mathrm{NV},i},
\nonumber\\
B_{\mathrm{DC},\perp}^{(i)}
 &=\sqrt{\lVert\boldsymbol B_{\mathrm{DC}}\rVert^2-
 \big(B_{\mathrm{DC},\parallel}^{(i)}\big)^2}.
\label{eq:nv_bias_mag}
\end{align}

The superscript $(i)$ identifies the NV orientation, not the magnet setting.
Below, the orientation index $i$ is omitted when a single selected NV axis
is being discussed; the corresponding local coordinates are denoted by
$(x',y',z')$.
The local transverse axes are chosen along the principal axes of the
strain/electric-field interaction when Eq.~\eqref{eq:nv_hamiltonian} is used.

For an individual NV orientation, the ground-state electron-spin Hamiltonian
in frequency units is
\begin{equation}
\frac{\hat H_{\mathrm{NV}}}{h}
 =D\hat S_{z'}^2
 +E_{x'}(\hat S_{x'}^2-\hat S_{y'}^2)
 +\gamma_e\boldsymbol B_{\mathrm{DC}}\cdot\hat{\boldsymbol S},
\label{eq:nv_hamiltonian}
\end{equation}
where $D\simeq2.87~\mathrm{GHz}$ is the zero-field splitting, $E_{x'}$ is the transverse strain/electric-field
interaction parameter, and $\gamma_e=g_e\mu_B/h\simeq28.0~\mathrm{MHz/mT}$ is the electron-spin
gyromagnetic ratio expressed in ordinary frequency units. \cite{Doherty2013}
The local $z'$ axis is aligned with the NV symmetry axis, while $x'$ and $y'$ are
chosen along the principal axes of the transverse strain/electric-field interaction.
This local coordinate system is distinct from the laboratory coordinate system
$(x,y,z)$ defined above.
The spin operators are dimensionless spin-one operators.

For NV orientations with a dominant longitudinal bias-field component,
the ODMR transition frequencies are approximately given by
$D\pm\gamma_e B^{(i)}_{\mathrm{DC},\parallel}$, provided that the effects of
transverse-field and strain-induced mixing are sufficiently weak.
For the selected NV orientation, the bias field is nearly perpendicular to the NV axis.
We denote the two upper spin eigenstates under this bias field by
$\ket{+}$ and $\ket{-}$.
When the transverse-field-induced admixture of the $\ket{0}$ basis
state is small, these eigenstates are approximately described by the symmetric
and antisymmetric superpositions of the $\ket{+1}$ and
$\ket{-1}$ spin states, respectively:
\begin{equation}
\ket{+}\simeq
 \frac{\ket{+1}+\ket{-1}}{\sqrt{2}},\qquad
\ket{-}\simeq
 \frac{\ket{+1}-\ket{-1}}{\sqrt{2}}.
\label{eq:upper_spin_states}
\end{equation}
Under the conditions of negligible residual longitudinal field, a transverse bias field aligned with the local strain principal axis, and a transverse Zeeman interaction much smaller than the zero-field splitting, the frequency separation between the $\ket{+}$ and $\ket{-}$ states can be approximated by a second-order expansion of the corresponding eigenenergy difference as described in Ref.~\onlinecite{Okaniwa2026arxiv}:
\begin{equation}
\nu_{+-}\simeq 2E_{x'}+
\frac{(\gamma_e B_{\mathrm{DC},\perp})^2}{D+E_{x'}}.
\label{eq}
\end{equation}
Thus, the RF resonance frequency can be tuned by varying the transverse
bias-field magnitude.
The RF frequency is chosen to be resonant with the
$\ket{+}$--$\ket{-}$ transition, while the MW
frequency probes spin transitions near the ground-state zero-field splitting.
For an arbitrary orientation of the transverse bias field relative to the
strain principal axes, a more complete diagonalization of the spin
Hamiltonian is required.

An RF magnetic field with a frequency $\nu_{\mathrm{RF}}$ close to $\nu_{+-}$
couples the two selected NV spin eigenstates through its component along the
selected NV axis. Throughout this work, RF magnetic-field amplitudes are quoted
as peak values. We denote the RF magnetic-field amplitude vector by
$\boldsymbol B_{\mathrm{RF}}(\boldsymbol r)$. The magnitude of its projection
along the $i$th NV axis is
\begin{equation}
B_{\mathrm{RF},\parallel}^{(i)}(\boldsymbol r)
=\left|\boldsymbol B_{\mathrm{RF}}(\boldsymbol r)
\cdot\boldsymbol n_{\mathrm{NV},i}\right|,
\label{eq:rf_projection}
\end{equation}
where $\boldsymbol n_{\mathrm{NV},i}$ is the unit vector along that axis.

When only one NV orientation is considered, the index $(i)$ is omitted.
Within the rotating-wave approximation, the splitting of each RF-dressed
resonance pair is approximately
\begin{equation}
\Delta\nu_{\mathrm{dress}}
 =\sqrt{(\nu_{+-}-\nu_{\mathrm{RF}})^2
 +(\gamma_e B_{\mathrm{RF},\parallel})^2}.
\label{eq:rf_dressed_splitting}
\end{equation}
More generally, the effective RF coupling contains a state-mixing factor $\cos\alpha$ induced by the transverse bias field. Here, $\alpha$ is the mixing parameter between the $\ket{+}$ and $\ket{0}$ spin states, determined by the transverse bias-field strength, zero-field splitting, and transverse strain interaction. \cite{Okaniwa2026arxiv}
This factor is close to unity under the present conditions, and is therefore
neglected in Eqs.~\eqref{eq:rf_dressed_splitting} and
\eqref{eq:resonant_splitting}.
At RF resonance, $\nu_{\mathrm{RF}}=\nu_{+-}$, the splitting reduces to
\begin{equation}
\Delta\nu_{\mathrm{dress}}\simeq\gamma_e B_{\mathrm{RF},\parallel}.
\label{eq:resonant_splitting}
\end{equation}
The coherent RF coupling splits the two corresponding MW ODMR transitions,
producing four resonances in the selected NV spectral region.

For the magnetic-field images, the RF magnetic-field projection amplitudes were obtained
using experimentally determined orientation-dependent calibration slopes
$\kappa_{\mathrm{exp}}^{(i)}$ and intercepts $b_{\mathrm{exp}}^{(i)}$:
\begin{equation}
B_{\mathrm{RF},\parallel}^{(i)}\simeq
 \frac{\Delta\nu_{\mathrm{dress}}^{(i)}-b_{\mathrm{exp}}^{(i)}}
 {\kappa_{\mathrm{exp}}^{(i)}},\qquad i=1,2,3.
\label{eq:field_conversion}
\end{equation}
The orientation-dependent calibration coefficients
were determined experimentally, as detailed in
Appendix~\ref{app:rf_calibration}.

\section{Experimental methods}
\label{sec:methods}

\subsection{Scanning NV microscope and probe}
The measurements were performed using a home-built scanning NV microscope
combining amplitude-modulation AFM (AM-AFM) and confocal
fluorescence detection, as schematically shown in Fig.~\ref{fig:setup}(a).
The scanning probe consisted of a quartz tuning-fork force sensor and a
tungsten tip to which an approximately $1~\mu\mathrm{m}$-diameter
single-crystal diamond particle containing an ensemble of NV centers was
attached using a UV-curable adhesive [Fig.~\ref{fig:setup}(b)].
MW excitation for ODMR was delivered through a coplanar waveguide connected to
a fine metallic wire positioned near the diamond particle, providing local
MW excitation while maintaining optical access to the sensing volume.

\subsection{Optical and MW instrumentation}
The NV centers were continuously excited with a 532-nm laser focused through an objective lens with a numerical aperture of 0.8. The excitation laser power measured before the objective was about $83~\mu\mathrm{W}$ during ODMR measurements.
The emitted fluorescence was separated from the excitation light by a dichroic mirror (DM), coupled into a single-mode optical fiber (SMF), and detected
using a single-photon-counting avalanche photodiode (APD).
A galvanometric mirror (GM) and a piezoelectric objective scanner were used to optimize the lateral alignment and axial focus of the excitation laser on the NV centers, respectively, based on their fluorescence intensity.
During magnetic-field imaging, tracking was performed after each scan line
rather than at every pixel.
A photoluminescence image of the diamond particle is shown in
Fig.~\ref{fig:setup}(c).

Continuous MW excitation was used to probe the NV spin transitions
through CW-ODMR, while an RF current applied to the sample circuit generated
the target AC magnetic field that induced spin dressing.
Thus, the MW field used for ODMR readout was distinct from the RF
magnetic field being measured.

\subsection{Bias-field control, sample circuit geometry, and coordinate system}
The static bias magnetic field was generated by a permanent magnet mounted on
two rotation stages and a translation stage.
Rotating the magnet allowed the bias-field direction to be adjusted nearly
perpendicular to a selected NV axis, enabling spectral selection of that
orientation.
Translating the magnet adjusted the bias-field magnitude and hence the
transition frequency between the $\ket{+}$ and $\ket{-}$
spin eigenstates.

\begin{figure*}[t]
\centering
\includegraphics[width=\textwidth]{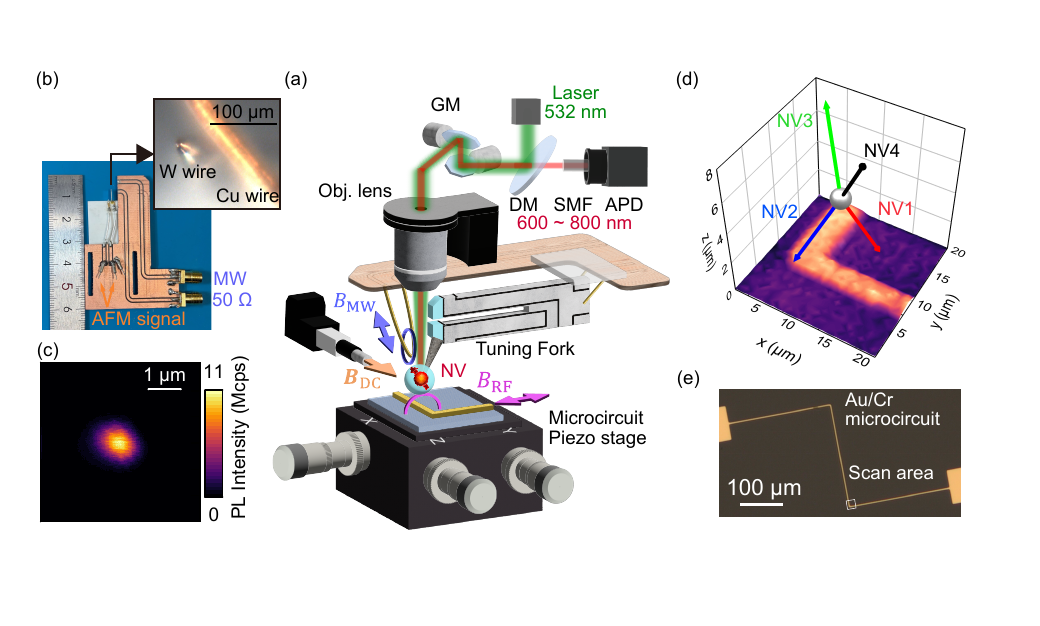}
\caption{Scanning NV microscope, probe, sample circuit, and coordinate system.
(a) Schematic of the scanning NV setup combining AM-AFM,
confocal fluorescence detection, MW excitation for CW-ODMR,
RF current excitation of the sample circuit, and piezoelectric
raster scanning of the sample. (b) Photograph of the assembled scanning probe with an
 approximately $1~\mu\mathrm{m}$-diameter single-crystal diamond particle containing an
 NV ensemble. (c) Photoluminescence (PL) image of the diamond particle at the probe
 tip. (d) AFM topography of the circuit with the four crystallographic NV-axis
 vectors (NV1--NV4) expressed in the laboratory coordinate system, for which
 $x$ and $y$ lie in the sample plane and $z$ is the sample normal. NV1--NV3 were
 selected successively for RF magnetic-field imaging. (e) Optical micrograph
 of the L-shaped Au/Cr microcircuit, with the scan area indicated.}
\label{fig:setup}
\end{figure*}

The sample consisted of an L-shaped Au/Cr microcircuit with a conductor width
of $2~\mu\mathrm{m}$ and a total metal thickness of $200~\mathrm{nm}$,
as shown in the optical micrograph in Fig.~\ref{fig:setup}(e).
Figure~\ref{fig:setup}(d) shows the AFM topography of the circuit together
with the NV-axis directions.
The laboratory coordinate system $(x,y,z)$ and the four NV-axis unit vectors
$\boldsymbol n_{\mathrm{NV},i}$ ($i=1,2,3,4$) are defined in
Sec.~\ref{sec:principle}.
The NV-axis directions used for the Biot--Savart calculations were
obtained from DC magnetic-field line-scan fitting, as detailed in
Appendix~\ref{app:axis_calibration}, and are listed in Table~\ref{tab:nv_axes}. The angular uncertainties of the determined NV-axis orientations
were evaluated by Monte Carlo analysis, yielding 68th-percentile
angular deviations of 0.59--0.75$^\circ$ for the four NV axes.

\begin{table}[t]
\caption{NV-axis unit-vector components in laboratory coordinates $(x,y,z)$.
The $i$th orientation is $\boldsymbol n_{\mathrm{NV},i}=
(n_{x,i},n_{y,i},n_{z,i})^{\mathsf{T}}$.
NV1--NV3 were selected successively for RF magnetic-field imaging;
NV4 is listed for completeness. Components are dimensionless.}
\label{tab:nv_axes}
\begin{ruledtabular}
\begin{tabular}{crrr}
Orientation & $n_{x,i}$ & $n_{y,i}$ & $n_{z,i}$ \\
\hline
NV1 & $0.470$ & $0.593$ & $-0.654$ \\
NV2 & $-0.539$ & $-0.681$ & $-0.495$ \\
NV3 & $-0.605$ & $0.551$ & $0.574$ \\
NV4 & $0.675$ & $-0.463$ & $0.575$ \\
\end{tabular}
\end{ruledtabular}
\end{table}

\subsection{CW-ESDR acquisition and magnetic-field imaging}
For imaging, an RF current at 33.5~MHz was applied to the L-shaped conductor to generate an AC Oersted magnetic field.
The sample circuit was mounted on a three-axis piezoelectric stage and raster-scanned relative to the probe, while AM-AFM feedback controlled the tip--sample spacing.
For each of the three selected NV orientations, the scanned area was $20\times20~\mu\mathrm{m}^2$ with $64\times64$ pixels.
At each pixel, a CW-ESDR spectrum was acquired by sweeping the MW frequency over a 35-MHz-wide range with 101 frequency points, while the laser and RF excitation were continuously applied. No pulsed spin control was required. The acquisition time was approximately 43~s per pixel.

The RF-excited spectrum contains four RF-dressed resonances in the selected NV region.
At each pixel, the higher-frequency RF-dressed resonance pair was fitted with a
sum of two Lorentzian functions, allowing their linewidths to vary independently.
The separation between the two fitted resonance centers was then converted
into $B_{\mathrm{RF},\parallel}^{(i)}$ using the corresponding experimentally
determined calibration slope $\kappa_{\mathrm{exp}}^{(i)}$ and intercept
$b_{\mathrm{exp}}^{(i)}$.
Pixels yielding negative RF magnetic-field projection amplitudes after calibration were
masked and excluded from the line-profile analysis.

The quantity $B_{\mathrm{RF},\parallel}^{(i)}$ is defined in
Eq.~\eqref{eq:rf_projection}. Only NV1--NV3 were used for the images reported
below. Because the measurement yields the magnitude of this projection, it does
not provide its sign or phase.

\section{Frequency-tunable and axis-selective RF sensing}
\label{sec:characterization}

\subsection{Selection of NV sensing axes}
We first characterized the orientation selectivity of the RF-dressed CW sensing scheme.
Figures~\ref{fig:characterization}(a)--(c) compare CW-ODMR spectra recorded without RF excitation and CW-ESDR spectra recorded with RF excitation for NV1, NV2, and NV3, respectively, selected by adjusting the static bias-field direction.
Without RF excitation, the full CW-ODMR spectrum contains eight resonances
associated with the four crystallographic NV orientations.
For each selected NV orientation, the two resonances correspond approximately to the $\ket{0}\leftrightarrow\ket{+}$ and $\ket{0}\leftrightarrow\ket{-}$ MW transitions, respectively. Here, $\ket{+}$ and $\ket{-}$ denote the two upper spin eigenstates under the nearly transverse bias field, while the lower eigenstate is predominantly $\ket{0}$. In contrast, the resonances from the nonselected NV orientations arise mainly from Zeeman-split transitions.
For each selected orientation, an RF field near resonance was applied to
couple the $\ket{+}$ and $\ket{-}$ states.
Under RF excitation, each of the corresponding MW transitions splits
into a pair of RF-dressed resonances, producing four resonance features in
the selected spectral region.
This behavior was observed for all three selected NV orientations,
demonstrating orientation-selective RF-dressed sensing without mechanically rotating the diamond probe or sample circuit.

In addition to the resonant splitting of the selected NV transitions,
additional dips appeared near the ODMR resonances of nonselected NV
orientations under RF excitation.
Their frequency offsets from the corresponding ODMR resonances were
approximately $\pm\nu_{\mathrm{RF}}$, resembling RF-induced
multifrequency resonances reported in previous studies. \cite{Onodera2024,Onodera2025}
These features may be associated with Floquet sidebands, although
their physical origin has not been conclusively established
in the present measurements.
These additional features were not used to determine the RF magnetic-field
projection magnitudes reported in this work.

\subsection{RF-induced splitting and RF-amplitude dependence}
Figure~\ref{fig:characterization}(d) shows an enlarged portion of the NV1
spectrum, in which two ODMR dips without RF excitation split into four
RF-dressed resonances under resonant RF driving.
This structure is consistent with coherent RF coupling between the
$\ket{+}$ and $\ket{-}$ spin eigenstates described in
Sec.~\ref{sec:principle}.

Figure~\ref{fig:characterization}(e) shows NV1 spectra as a function of the
estimated RF magnetic-field projection amplitude $B_{\mathrm{RF},\parallel}^{(1)}$,
ranging from zero to approximately $600~\mu\mathrm{T}$.
The spectra include an RF-off CW-ODMR reference spectrum and RF-on CW-ESDR
spectra measured at different RF magnetic-field projection amplitudes.
The RF drive was adjusted near the $\ket{+}$--$\ket{-}$
transition estimated from the RF-off spectrum.
The separation between the dressed-state resonances increases with the applied
projection amplitude, consistent with increasing spin--RF coupling.
For spatial magnetic-field imaging, the local RF magnetic-field projection
magnitudes were calculated from the measured dressed-state splittings using
the empirical calibration of Eq.~\eqref{eq:field_conversion}.
The observed amplitude dependence does not, by itself, establish the lower or
upper limits of the quantitative dynamic range.

\subsection{Bias-field control of the detection frequency}

\begin{figure*}[t]
\centering
\includegraphics[width=\textwidth]{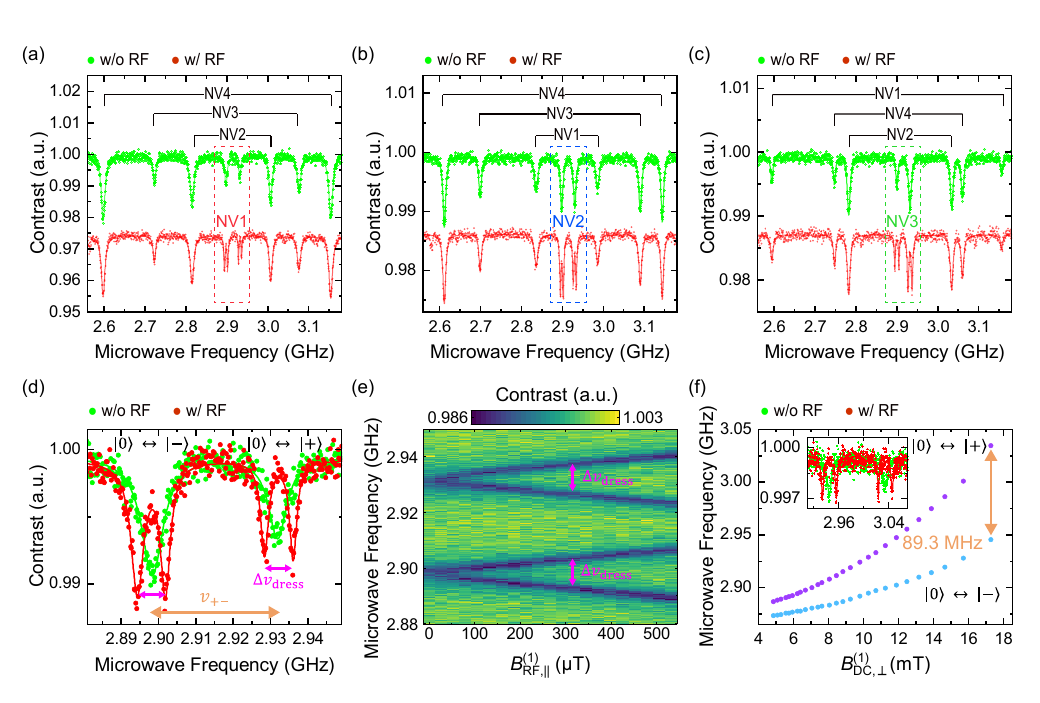}
\caption{Axis-selective and frequency-tunable RF-dressed continuous-wave
magnetometry. (a)--(c) CW-ODMR spectra without RF excitation and CW-ESDR
spectra with RF excitation, measured under static bias-field configurations
nearly perpendicular to NV1, NV2, and NV3, respectively. Each spectrum
includes resonances from all four NV crystallographic orientations.
The selected NV resonances exhibit RF-induced splitting. Additional dips
are visible near nonselected NV resonances.
(d) Enlarged view of the NV1 resonance region from (a), showing two
CW-ODMR resonances split into four RF-dressed resonances under RF excitation.
(e) NV1 spectra as a function of the estimated RF magnetic-field projection
amplitude $B_{\mathrm{RF},\parallel}^{(1)}$ (up to approximately $600~\mu\mathrm{T}$),
including the RF-off CW-ODMR reference and RF-on CW-ESDR spectra.
(f) Frequencies of the two selected NV1 CW-ODMR resonances as a function of the transverse bias-field component estimated by Hamiltonian fitting. Their separation, $\nu_{+-}$, increases from 14.4 to 89.3~MHz. The inset shows RF-dressed splitting at an RF frequency of 89.3~MHz.}

\label{fig:characterization}
\end{figure*}

Figure~\ref{fig:characterization}(f) shows the two CW-ODMR transition
frequencies of NV1 as a function of the transverse bias-field component
$B_{\mathrm{DC},\perp}^{(1)}$. The magnet orientation was kept fixed while
its position was adjusted to vary the bias-field strength. The transverse field component was estimated from the two measured ODMR transition frequencies using the Hamiltonian-fitting procedure described in Appendix~\ref{app:transverse_bias}.
The transition frequency $\nu_{+-}$ was
obtained from the separation between these resonances, yielding a tuning range
from 14.4 to 89.3~MHz.
The tuning curve therefore represents ODMR-derived
frequencies rather than direct RF-dressed measurements at
every frequency.
The same panel includes a representative RF-off CW-ODMR spectrum and an RF-on CW-ESDR spectrum at 89.3~MHz, where two selected resonances become four RF-dressed resonances.

The observed ODMR contrast decreased toward higher detection frequencies as the
transverse bias field was increased.
This observation is consistent with changes in optical spin polarization and
fluorescence readout arising from transverse-field-induced mixing of the
spin-projection basis states.\cite{Tetienne2012,Okaniwa2026arxiv}
Thus, frequency tunability and spectral readout quality must be considered
separately when assessing the practical operating range.

\section{Axis-selective AC magnetic-field imaging}
\label{sec:imaging}

\subsection{Imaging with three selected NV orientations}
We next demonstrated spatially resolved RF magnetic-field imaging at
33.5~MHz using RF-dressed CW SNVM.
Figures~\ref{fig:imaging}(a)--(c) show the measured RF magnetic-field
projection magnitudes
$B_{\mathrm{RF},\parallel}^{(i)}(\boldsymbol r)$ for NV1, NV2, and NV3
($i=1,2,3$), respectively.
The selected NV axis was changed by adjusting the static bias-field
direction without mechanically rotating the probe or sample circuit.

The resulting images exhibit distinct spatial distributions despite
being acquired from the same current-carrying structure.
Because the direction of the local Oersted field varies across the
L-shaped conductor, the RF magnetic-field projection magnitude exhibits
a different spatial dependence for each selected NV axis, as expressed
in Eq.~\eqref{eq:rf_projection}.

\subsection{Comparison with Biot--Savart calculations}

To interpret the distinct spatial distributions observed for the three
selected NV orientations, we performed Biot--Savart calculations of the
Oersted field generated by the finite-width L-shaped conductor and evaluated
its projection magnitude along each NV axis. The conductor model used a width
of $2~\mu\mathrm{m}$ and a thickness of $200~\mathrm{nm}$.

In the Biot--Savart calculation, the reference height of the NV-ensemble
sphere center was fixed at $h_{\mathrm{NV}}=0.70~\mu\mathrm{m}$, and the
RF current amplitude was fixed at $I_{\mathrm{RF}}=4.65~\mathrm{mA}$.
These parameters were determined independently from magnetic-field line-scan
analyses, as described in Appendix~\ref{app:axis_calibration}. The same RF
current amplitude and sensing height were used for all three calculated maps;
no map-specific amplitude scaling or height adjustment was applied.

The calculated projection magnitudes were
\begin{equation}
B_{\mathrm{RF},\parallel}^{(i),\mathrm{calc}}(\boldsymbol r)
=\left|\boldsymbol B_{\mathrm{RF}}^{\mathrm{calc}}
(\boldsymbol r;I_{\mathrm{RF}},h_{\mathrm{NV}})
\cdot\boldsymbol n_{\mathrm{NV},i}\right|,
\quad i=1,2,3.
\label{eq:rf_projection_calc}
\end{equation}

Figures~\ref{fig:imaging}(d)--(f) show the calculated maps for NV1, NV2, and
NV3, respectively. The calculated projection maps qualitatively reproduce the
spatial patterns observed experimentally, supporting the consistency of the
measured RF magnetic-field projection maps with the Biot--Savart model.

\subsection{Magnetic-field line profiles}

\begin{figure*}[t]
\centering
\includegraphics[width=\textwidth]{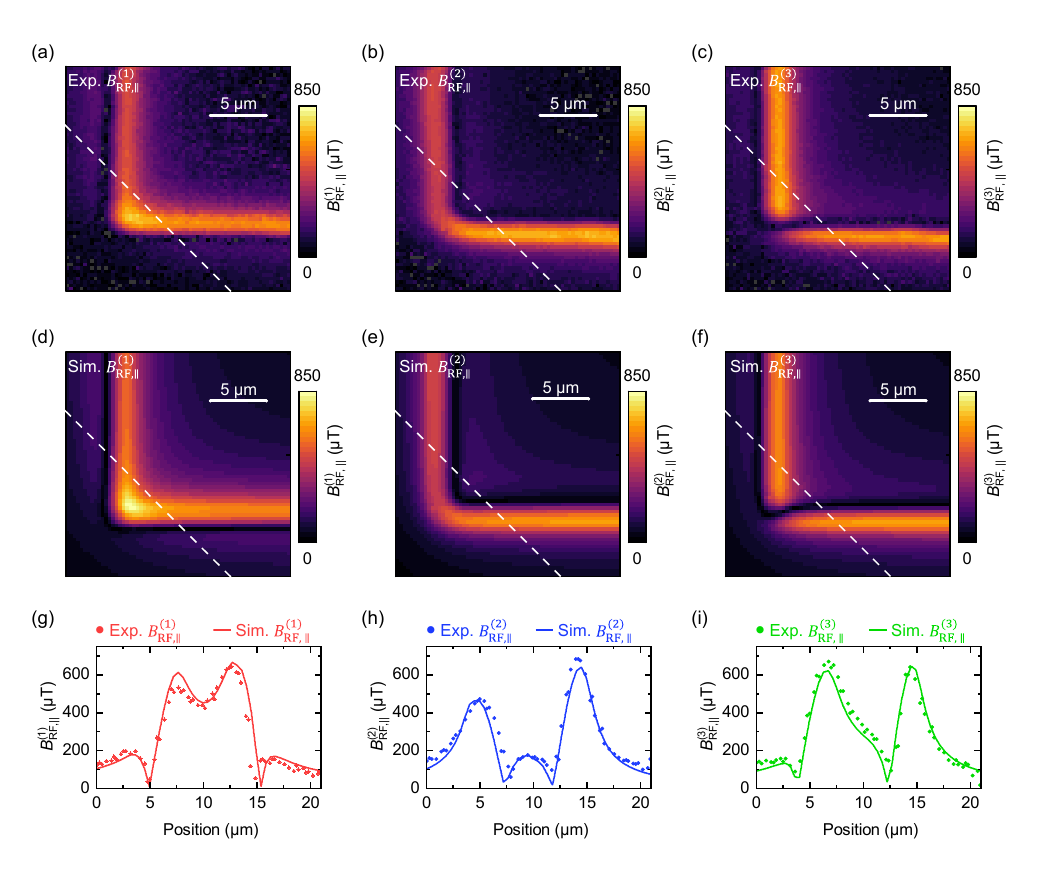}
\caption{
Axis-selective AC magnetic-field imaging at 33.5~MHz using
RF-dressed CW SNVM.
(a)--(c) Measured RF magnetic-field projection magnitudes
$B_{\mathrm{RF},\parallel}^{(i)}$ for the successively
selected NV1, NV2, and NV3 orientations, respectively.
The amplitudes were obtained from the splitting of the
higher-frequency RF-dressed resonance pair using the
experimentally calibrated slopes and intercepts.
Negative calibrated amplitudes were masked.
Each map covers $20\times20~\mu\mathrm{m}^{2}$ with
$64\times64$ pixels.
(d)--(f) Corresponding calculated projection magnitudes
for NV1, NV2, and NV3, respectively, obtained from a
finite-cross-section Biot--Savart model.
The reference height of the modeled diamond-particle center
was fixed at $h_{\mathrm{NV}}=0.70~\mu\mathrm{m}$,
as determined from the DC magnetic-field analysis,
and the independently determined RF current amplitude was
$I_{\mathrm{RF}}=4.65~\mathrm{mA}$.
(g)--(i) Experimental (symbols) and calculated (solid lines)
line profiles for NV1, NV2, and NV3, respectively,
extracted along the white dashed lines in the corresponding maps.
Masked pixels were excluded from the experimental profiles.
}
\label{fig:imaging}
\end{figure*}

Figures~\ref{fig:imaging}(g)--(i) compare the experimental (symbols) and calculated (solid lines) RF magnetic-field projection profiles for NV1, NV2, and NV3, respectively. The profiles were extracted along the white dashed lines in the corresponding measured and calculated maps. Masked pixels were excluded from the experimental profiles.

Because the line crosses both arms of the L-shaped conductor, the direction of the local Oersted field varies along the scan path. Consequently, the three selected NV orientations exhibit distinct profile shapes, including different relative heights of the two major peaks and different depths of the intervening minima. These differences cannot be described by a simple rescaling of the overall signal amplitude. Instead, they reflect the spatially varying projection of the same local RF magnetic field onto the three NV axes. The corresponding Biot--Savart projections reproduce the positions of the major peaks and minima as well as the orientation-dependent changes in their relative amplitudes. This agreement supports the interpretation that the measurements selectively probe different projections of the local RF magnetic field.

Quantitative deviations nevertheless remain in both weak- and strong-RF-field regions. In weak-field regions, finite experimental amplitudes remain at some positions where the calculated field projection becomes small, whereas larger absolute discrepancies are observed in the high-field regions. Possible origins of these deviations are discussed in Sec.~\ref{sec:discussion}.

\section{Discussion}
\label{sec:discussion}

The three NV-axis measurements obtained in this work provide
complementary spatial information about the local RF magnetic field. Because the present measurement yields only projection magnitudes, full vector reconstruction additionally requires relative sign or phase information for the measured projections.
In a future phase-sensitive implementation, measuring the phase of each RF response relative to a known RF reference could provide the complex projection amplitudes required to reconstruct the RF magnetic-field vector from measurements along multiple NV orientations.

The quantitative accuracy of the RF magnetic-field projection amplitudes depends on the empirical calibration of the RF-dressed splitting. For each NV orientation, a linear calibration was obtained from the higher-frequency CW-ESDR resonance pair using only splittings exceeding the mean full width at half maximum (FWHM) of the two RF-dressed resonance dips, with both the slope and intercept treated as free parameters. The resulting slopes ranged from 25.2 to 27.3~MHz/mT, while the nonzero intercepts corresponded to equivalent field offsets of approximately 59--67~$\mu$T. Although the microscopic origin of these intercepts and the differences among the orientation-dependent slopes have not yet been established, the experimentally determined calibration coefficients provide a practical empirical conversion from the measured splittings to RF magnetic-field projection amplitudes.

The line-profile comparison reveals a characteristic limitation in weak-field regions. When the RF-dressed splitting becomes comparable to or smaller than the resonance linewidth, the two resonances can no longer be reliably resolved, and fitting them with two Lorentzian functions can yield an apparent nonzero splitting even when the actual splitting is small. This effect, together with the empirical calibration intercept, can lead to an overestimation of the RF magnetic-field projection amplitude in weak-field regions. In the lowest-field regions, the fitted splitting can fall below the empirical calibration intercept, yielding a negative calibrated amplitude; these pixels were therefore masked and excluded from the quantitative comparison. Improved spectral resolution using higher-quality diamond probes and narrower RF-dressed resonances, together with increased spectral contrast and fluorescence collection efficiency, should improve the accuracy of weak-field measurements. \cite{Maletinsky2012,Tabuchi2023,Beaver2024}

In strong-RF-field regions, the spatial magnetic-field gradient is large, so small uncertainties in the probe position, sensing height, and assumed current distribution can produce relatively large changes in the calculated field. These effects can therefore contribute to the larger absolute deviations observed between experiment and calculation in these regions. The approximately $1~\mu\mathrm{m}$-diameter diamond particle used in the present experiment also introduces spatial averaging over a finite sensing volume, making the calculated profiles sensitive to assumptions about the particle geometry and NV distribution. Using a smaller diamond probe would reduce this spatial averaging and make the sensor more closely approximate a point-like magnetic-field probe. This could simultaneously improve spatial resolution and reduce model-related uncertainty in the quantitative comparison between experiment and calculation. Ultimately, single-NV probes provide a route toward nanoscale RF magnetic-field imaging. \cite{Maletinsky2012} Faster imaging could also be achieved by estimating the RF-dressed splitting from a limited number of MW frequency points rather than acquiring a complete CW-ESDR spectrum at every pixel.

At still higher RF amplitudes, an additional limitation arises from the spectroscopy itself. The approximately linear relationship between RF-dressed splitting and RF magnetic-field projection amplitude used in the present calibration may break down when the RF coupling strength becomes comparable to the $\ket{+}$--$\ket{-}$ transition frequency $\nu_{+-}$. In this regime, deviations from the rotating-wave approximation and overlapping spectral features can complicate field estimation from a single resonance pair. Rather than representing only a limitation, however, the strong-driving regime may provide additional sensing opportunities. Floquet-based analysis has been shown to reproduce ESDR spectra beyond the rotating-wave approximation and may enable quantitative RF-field estimation over a wider dynamic range. \cite{Mikawa2023} Increasing $\nu_{+-}$ through a stronger transverse bias field could also shift the onset of this nonlinear regime toward higher RF amplitudes, although the associated change in ODMR contrast must be considered.

Strong RF driving can additionally generate multifrequency resonances that provide information beyond the primary RF-dressed resonance pair. Such resonances have been used for RF-amplitude calibration and RF magnetic-field imaging. \cite{Onodera2024,Onodera2025} In the present measurements, additional RF-induced dips were also observed near the resonances of nonselected NV orientations. Their microscopic origin and quantitative relation to the RF-field components have not yet been established; however, if these features can be quantitatively described within a Floquet or multifrequency-resonance framework, they could provide simultaneous access to responses associated with multiple NV orientations under a single bias-field configuration. Such an approach could reduce the need to sequentially change the bias-field direction for each selected NV axis and thereby enable faster multi-axis RF magnetic-field imaging.

Extending the detection frequency provides another promising direction for the present method. Transverse bias-field control enabled the $\ket{+}$--$\ket{-}$ transition frequency to be tuned from 14.4 to 89.3~MHz, as inferred from CW-ODMR measurements and Hamiltonian analysis, with RF-dressed resonances directly observed at 89.3~MHz. The upper end of this range was limited by the available travel of the bias magnet rather than by an established fundamental limit of the sensing principle. Furthermore, RF magnetic-field detection up to 632~MHz has been demonstrated using the same RF-dressed CW sensing principle in a confocal NV setup. \cite{Okaniwa2026arxiv} Combining higher-quality and smaller scanning probes, Floquet- or multifrequency-resonance-based analysis, and phase-sensitive detection referenced to a known RF signal could therefore extend RF-dressed CW SNVM toward higher spatial resolution, a broader dynamic range, and multi-axis or vector-resolved AC magnetic-field imaging in the hundreds-of-megahertz regime.

\section{Conclusion}
\label{sec:conclusion}

We demonstrated axis-selective AC magnetic-field imaging at 33.5~MHz using RF-dressed CW SNVM, extending AC magnetic-field imaging with SNVM into the tens-of-megahertz regime.
The CW approach avoids short, high-power MW control pulses and
synchronization between the RF field and a pulsed sensing sequence.
By successively selecting three crystallographic NV orientations through
transverse bias-field control, we obtained spatial maps of the RF
magnetic-field projection magnitudes from the same L-shaped microcircuit.
The calculated projection maps reproduce the main spatial
patterns observed experimentally, supporting their consistency with
the Biot--Savart model.
RF-dressed resonances were experimentally observed at frequencies up to
89.3~MHz, while spatially resolved AC magnetic-field imaging was
demonstrated at 33.5~MHz.
The demonstrated axis-selective imaging capability provides a foundation
for future vector-resolved AC magnetic-field imaging.

\begin{acknowledgments}
We are grateful to Dr. Ryo Tamaki for helpful discussions
and suggestions that greatly improved our work.

This work was supported by the MEXT Quantum Leap Flagship Program (Q-LEAP) (Grant No. JPMXS0118067395), JSPS KAKENHI (Grant No. JP25K01292), JSPS Grant-in-Aid for JSPS Fellows (Grant No. JP24KJ1964), SIP Program, the Center for Spintronics Research Network (CSRN), Keio University, the Program for the Advancement of Next Generation Research Projects, Keio University, and JST CREST (Grant No. JPMJCR24A5).
\end{acknowledgments}

\section*{AUTHOR DECLARATIONS}

\subsection*{Conflict of Interest}

The authors have no conflicts to disclose.

\subsection*{Author Contributions}

\textbf{Shuntaro Komuro:}
Conceptualization (lead); data curation (lead); formal analysis (lead); investigation (lead); methodology (lead); project administration (supporting); software (supporting); visualization (lead); writing -- original draft (lead); writing -- review and editing (equal). \textbf{Ryusei Okaniwa:}
Conceptualization (equal); funding acquisition (supporting); formal analysis (supporting); investigation (supporting); methodology (equal); software (equal); validation (supporting); writing -- review and editing (equal). \textbf{Kazuma Okura:} Conceptualization (equal); investigation (supporting); resources (equal); project administration (supporting). \textbf{Kunitaka Hayashi:}
Resources (supporting); software (lead); writing -- review and editing (supporting). \textbf{Toshu An:}
Conceptualization (equal); methodology (supporting); project administration (equal); resources (supporting); supervision (equal); writing -- review and editing (supporting). \textbf{Junko Ishi-Hayase:}
Conceptualization (equal); funding acquisition (lead);
investigation (supporting); methodology (supporting);
project administration (lead); resources (lead);
supervision (lead); writing -- original draft (supporting); writing -- review and editing (lead).

\section*{DATA AVAILABILITY STATEMENT}

The data that support the findings of this study are available
from the corresponding authors upon reasonable request.

\appendix
\section{Independent determination of parameters used in the magnetic-field calculations}
\label{app:axis_calibration}

The magnetic-field calculations in Fig.~\ref{fig:imaging}(d)--(f) require three experimentally determined quantities:
(i) the orientations of the four crystallographic NV axes in the laboratory coordinate system, $\boldsymbol n_{\mathrm{NV},i}$;
(ii) the absolute NV sensing height, $h_{\mathrm{NV}}$; and
(iii) the RF current amplitude, $I_{\mathrm{RF}}$, flowing through the sample circuit at 33.5~MHz.
Quantities (i) and (ii) were determined jointly from the DC magnetic-field line-scan data, as described in Secs.~\ref{app:dc_reconstruction}--\ref{app:global_fit}, and the uncertainty in the NV-axis orientations was evaluated as described in Sec.~\ref{app:axis_uncertainty}.
Quantity (iii) was subsequently determined from an independent
33.5-MHz RF magnetic-field line scan, with the sensing height
and NV-axis orientations obtained from the DC analysis and the
sample circuit geometry held fixed, as described in Sec.~\ref{app:rf_current}.
The RF field-conversion coefficients required for the RF line-scan analysis were independently calibrated as described in Appendix~\ref{app:rf_calibration}.
The transverse bias-field component used for Fig.~\ref{fig:characterization}(f), which is separate from the parameter determination for Fig.~\ref{fig:imaging}, is described in Appendix~\ref{app:transverse_bias}.

The DC and RF line scans were acquired across a straight section of the same L-shaped sample circuit used for the two-dimensional RF magnetic-field imaging.
The line-scan region was sufficiently far from the bend that the conductor could be approximated as straight.

\subsection{Reconstruction of the total static magnetic-field vector from DC line scans}
\label{app:dc_reconstruction}

CW-ODMR line scans acquired under positive and negative DC voltages were used to reconstruct the three-dimensional total static magnetic-field vector at each scan position.
Positive and negative voltages of equal magnitude and opposite polarity, $+V_{\mathrm{DC}}$ and $-V_{\mathrm{DC}}$, were applied to the sample circuit.
At each position of the line scan, the eight CW-ODMR resonances originating from the four NV orientations were fitted with Lorentzian functions to determine their resonance frequencies.

The total static magnetic fields experienced by the NV centers under the positive and negative DC voltages are written as
\begin{equation}
\boldsymbol B_{\mathrm{tot}}^{\pm}(x)
=
\boldsymbol B_{\mathrm{bias}}
+
\boldsymbol B_{\mathrm{bg}}(x)
\pm
\boldsymbol B_{\mathrm{circ}}(x),
\label{eq:app_total_field}
\end{equation}
where $\boldsymbol B_{\mathrm{bias}}$ is the static bias field applied by the permanent magnet,
$\boldsymbol B_{\mathrm{bg}}(x)$ represents voltage-independent static background fields common to the positive- and negative-voltage measurements,
and $\boldsymbol B_{\mathrm{circ}}(x)$ is the magnetic field generated by the DC current flowing through the sample circuit.
For equal current amplitudes with opposite directions under $+V_{\mathrm{DC}}$ and $-V_{\mathrm{DC}}$, only the sign of $\boldsymbol B_{\mathrm{circ}}(x)$ is reversed.

For each NV orientation, the measured CW-ODMR resonance frequencies were analyzed using the spin-1 Hamiltonian
\begin{align}
\frac{\hat H^{\pm}(x)}{h}
={}&
D\hat S_{z'}^{2}
+
E_{x'}
\left(
\hat S_{x'}^{2}
-
\hat S_{y'}^{2}
\right)
\nonumber+
E_{y'}
\left(
\hat S_{x'}\hat S_{y'}
+
\hat S_{y'}\hat S_{x'}
\right)
\nonumber\\
&+
\gamma_e
\left[
\boldsymbol B_{\mathrm{bias}}
+
\boldsymbol B_{\mathrm{bg}}(x)
\pm
\boldsymbol B_{\mathrm{circ}}(x)
\right]
\cdot
\hat{\boldsymbol S},
\label{eq:app_hamiltonian}
\end{align}
where $z'$ is parallel to the corresponding NV axis and $x'$ and $y'$ define the transverse directions.
Unlike the principal-axis form used in Eq.~(2), the general Hamiltonian used for the magnetic-field reconstruction retains both transverse strain/electric-field interaction parameters, $E_{x'}$ and $E_{y'}$.
Here, $D\simeq2.870~\mathrm{GHz}$ is the zero-field splitting,
$\gamma_e\simeq28.0~\mathrm{MHz/mT}$ is the electron-spin gyromagnetic ratio used for the DC magnetic-field reconstruction,
and $E_{x'}$ and $E_{y'}$ represent transverse strain/electric-field interactions. For the magnetic-field reconstruction, a Cartesian coordinate system fixed to the diamond crystal was introduced, in which the four crystallographic NV axes have fixed orientations.
For each NV orientation, the total magnetic-field vector was then transformed into the corresponding local NV coordinate system before diagonalization.

To account for inhomogeneous transverse strain/electric-field interactions within the NV ensemble,
$E_{x'}$ and $E_{y'}$ were independently sampled from zero-mean Gaussian distributions with a standard deviation of 3.30~MHz.
A total of 128 strain realizations, corresponding to 128 pairs of $(E_{x'},E_{y'})$, were used.
For each trial three-dimensional magnetic-field vector, the Hamiltonian was diagonalized for all four NV orientations and all strain realizations.
The two transition frequencies for each NV orientation were calculated for each realization and then averaged over the strain distribution, yielding eight strain-averaged predicted CW-ODMR transition frequencies.

The three-dimensional total magnetic-field vector was obtained by minimizing the deviations between these eight predicted and measured transition frequencies at each scan position.
This reconstruction therefore accounts for transverse magnetic-field components and the inhomogeneous strain/electric-field distribution rather than relying on a simple Zeeman-splitting approximation.
This reconstruction procedure follows the approach reported in Ref.~\onlinecite{Okaniwa2026arxiv}.
The analysis yielded the three-dimensional total magnetic-field vectors
$\boldsymbol B_{\mathrm{tot}}^{+}(x)$ and
$\boldsymbol B_{\mathrm{tot}}^{-}(x)$
at each position of the positive- and negative-voltage line scans, respectively.

\subsection{Extraction of the DC magnetic field generated by the sample circuit current}
\label{app:dc_extraction}

The magnetic field generated by the sample circuit current was isolated from the reconstructed total static field by taking the component antisymmetric with respect to reversal of the applied DC voltage:
\begin{equation}
\boldsymbol B_{\mathrm{circ}}(x)
=
\frac{
\boldsymbol B_{\mathrm{tot}}^{+}(x)
-
\boldsymbol B_{\mathrm{tot}}^{-}(x)
}{2}.
\label{eq:app_dc_field}
\end{equation}

Using Eq.~\eqref{eq:app_total_field},
the applied bias field $\boldsymbol B_{\mathrm{bias}}$ and the voltage-independent background field $\boldsymbol B_{\mathrm{bg}}(x)$ cancel in this difference, leaving the experimentally reconstructed three-dimensional magnetic field generated by the DC current flowing through the sample circuit.
The resulting vector profile $\boldsymbol B_{\mathrm{circ}}(x)$ was used for the global Biot--Savart fit described below.

\subsection{Global Biot--Savart fit for the NV-axis orientations and sensing height}
\label{app:global_fit}

The reconstructed circuit-generated magnetic-field profile was globally fitted to a finite-cross-section Biot--Savart model to determine quantities (i) and (ii) defined above.
The straight section of the sample circuit was modeled as a conductor with a width of $w=2~\mu\mathrm{m}$ and a thickness of $t=0.20~\mu\mathrm{m}$, with a uniform current density over its cross section.
For the line-scan geometry, the straight conductor and current direction were along the laboratory $y$ axis, while the scan direction was along the laboratory $x$ axis.
The three-dimensional magnetic field per unit current, $\boldsymbol b_{\mathrm{BS}}$, was calculated in the laboratory coordinate system $(x,y,z)$ from the Biot--Savart law.

The AFM measurement provides the relative change in probe height as the probe follows the sample circuit topography,
but does not independently determine the absolute distance between the diamond particle and the substrate.
The height of the center of the diamond particle was therefore written as
\begin{equation}
z_{\mathrm{center}}(x)
=
h_0
+
h_{\mathrm{AFM}}(x)
+
R_{\mathrm{DP}},
\label{eq:app_height}
\end{equation}
where $h_0$ is the reference separation between the planar substrate surface outside the conductor and the bottom of the diamond particle,
$h_{\mathrm{AFM}}(x)$ is the relative AFM height profile,
and $R_{\mathrm{DP}}$ is the radius of the diamond particle.

Because the approximately $1~\mu\mathrm{m}$-diameter diamond particle contains an NV ensemble distributed over a finite sensing volume,
spatial averaging of the calculated magnetic field was included in the model.
The diamond particle was approximated as a sphere with
$R_{\mathrm{DP}}=0.50~\mu\mathrm{m}$,
and the NV centers were assumed to be uniformly distributed within the sphere.
The Cartesian components of the calculated magnetic-field vector were averaged over the sensing volume while retaining their signs:
\begin{equation}
\boldsymbol b_{\mathrm{avg}}(x;h_0)
=
\frac{1}{V_{\mathrm{DP}}}
\int_{V_{\mathrm{DP}}(x;h_0)}
\boldsymbol b_{\mathrm{BS}}(\boldsymbol r)\,dV.
\label{eq:app_volume_average}
\end{equation}
Here, $\boldsymbol b_{\mathrm{avg}}$ is expressed in the laboratory coordinate system.

The reconstructed magnetic-field vectors obtained from the CW-ODMR analysis are expressed in the diamond coordinate system, whereas
$\boldsymbol b_{\mathrm{avg}}$ is calculated in the laboratory coordinate system.
Their relationship was therefore described by a rotation matrix $\mathcal R$ defined as the rotation from the laboratory coordinate system to the diamond coordinate system:
\begin{equation}
\boldsymbol B_{\mathrm{circ}}(x_j)
\simeq
I_{\mathrm{DC}}
\mathcal R
\boldsymbol b_{\mathrm{avg}}(x_j;h_0),
\label{eq:app_rotation}
\end{equation}
where $x_j$ denotes the position of the $j$th sampling point along the line scan, and $I_{\mathrm{DC}}$ is the DC current amplitude.

The free parameters of the global fit were the reference separation
$h_0$ [equivalently, the center-referenced sensing height
$h_{\mathrm{NV}}=h_0+R_{\mathrm{DP}}$],
the rotation matrix $\mathcal R$, and the DC current amplitude
$I_{\mathrm{DC}}$.
They were determined by minimizing
\begin{equation}
\chi^2(h_0,\mathcal R,I_{\mathrm{DC}})
=
\sum_j
\left\|
\boldsymbol B_{\mathrm{circ}}(x_j)
-
I_{\mathrm{DC}}
\mathcal R
\boldsymbol b_{\mathrm{avg}}(x_j;h_0)
\right\|^2.
\label{eq:app_cost}
\end{equation}

For each trial sensing height,
the rotation matrix $\mathcal R$ that best aligned the calculated and reconstructed three-dimensional magnetic-field vectors was obtained using singular-value decomposition,
with the constraint $\det\mathcal R=+1$,
and $I_{\mathrm{DC}}$ was determined by least-squares optimization.

The fit yielded a reference separation of
$h_0=0.20~\mu\mathrm{m}$,
corresponding to an effective sensing height referenced to the center of the spherical diamond particle of
$h_{\mathrm{NV}}=h_0+R_{\mathrm{DP}}=0.70~\mu\mathrm{m}$.

The four NV crystallographic axes in the diamond coordinate system,
$\boldsymbol n_{\mathrm{NV},i}^{\mathrm{diamond}}$,
are fixed by the diamond crystal structure.
Once the fitted rotation matrix $\mathcal R$ is obtained,
the NV-axis orientations in the laboratory coordinate system are determined directly from
\begin{equation}
\boldsymbol n_{\mathrm{NV},i}
=
\mathcal R^{T}
\boldsymbol n_{\mathrm{NV},i}^{\mathrm{diamond}},
\qquad
i=1,2,3,4.
\label{eq:app_NVrotation}
\end{equation}

Thus, the Cartesian components of the four NV-axis vectors were not fitted independently.
Instead, all four NV-axis orientations were determined from the single fitted rotation matrix $\mathcal R$.
The resulting NV-axis vectors are listed in Table~I.

\subsection{Uncertainty in the NV-axis orientations}
\label{app:axis_uncertainty}

The uncertainty in the NV-axis orientations obtained from the global DC line-scan fit was evaluated using 100 Monte Carlo trials.
This analysis quantifies how uncertainties in the reconstructed magnetic-field profile and in the geometrical parameters of the Biot--Savart model propagate into the fitted rotation matrix and the resulting NV-axis orientations.

In each Monte Carlo trial,
the residuals of the reconstructed magnetic-field profile were resampled,
while the conductor width, conductor thickness, and the overall line-scan position relative to the conductor were varied according to their assigned uncertainties.
The global Biot--Savart fitting procedure was then repeated,
yielding a new rotation matrix $\mathcal R^{(k)}$ and the corresponding four NV-axis orientations
$\boldsymbol n_{\mathrm{NV},i}^{(k)}$ for each trial.

For NV orientation $i$,
the angular deviation between the best-fit direction
$\boldsymbol n_{\mathrm{NV},i}^{(0)}$
and the direction obtained in Monte Carlo trial $k$,
$\boldsymbol n_{\mathrm{NV},i}^{(k)}$,
was defined as
\begin{equation}
\delta\theta_i^{(k)}
=
\cos^{-1}
\left[
\boldsymbol n_{\mathrm{NV},i}^{(0)}
\cdot
\boldsymbol n_{\mathrm{NV},i}^{(k)}
\right].
\label{eq:app_axis_uncertainty}
\end{equation}

The 68th-percentile angular deviations were
$0.69^\circ$, $0.59^\circ$, $0.60^\circ$, and $0.75^\circ$
for NV1, NV2, NV3, and NV4, respectively.
The corresponding 95th-percentile values were
$1.14^\circ$, $0.94^\circ$, $0.99^\circ$, and $1.19^\circ$,
respectively.
These values quantify the stability of the inferred NV-axis orientations against the measurement and geometrical uncertainties included in the DC magnetic-field and Biot--Savart analysis.

\subsection{Determination of the RF current amplitude from a 33.5-MHz RF magnetic-field line scan}
\label{app:rf_current}

Quantity (iii), the RF current amplitude $I_{\mathrm{RF}}$, was determined independently from a separate RF magnetic-field line scan acquired at 33.5~MHz across a straight section of the same L-shaped sample circuit.

At each position,
the measured splitting of the higher-frequency RF-dressed resonance pair was converted into the RF magnetic-field projection amplitude along the selected NV axis using the orientation-dependent empirical RF field-conversion coefficients determined in Appendix~\ref{app:rf_calibration}.

In the Biot--Savart calculation of the RF line scan,
the sample circuit geometry was fixed, and the sensing height
$h_{\mathrm{NV}}$ and NV-axis orientations
$\boldsymbol n_{\mathrm{NV},i}$ determined from the DC analysis
were held fixed.
The only fitted parameter was therefore the RF current amplitude $I_{\mathrm{RF}}$.
The calculated RF magnetic field was projected onto the corresponding NV axis and compared with the experimentally determined RF magnetic-field projection profile.

Minimization of the difference between the measured and calculated RF magnetic-field profiles yielded
$I_{\mathrm{RF}}=4.65~\mathrm{mA}$,
where $I_{\mathrm{RF}}$ denotes the peak RF current amplitude.

The same value of $I_{\mathrm{RF}}$,
together with the same sensing height $h_{\mathrm{NV}}$,
NV-axis orientations $\boldsymbol n_{\mathrm{NV},i}$,
and sample circuit geometry,
was subsequently used for the NV1, NV2, and NV3 calculations in Fig.~\ref{fig:imaging}(d)--(f).
Thus, all three calculated maps were generated using a common set of independently determined geometrical, orientational, and electrical parameters,
rather than by individually adjusting the current amplitude, sensing height, or NV-axis orientation to reproduce each measured image.

\section{Experimental calibration of RF field-conversion coefficients}
\label{app:rf_calibration}

This calibration establishes the experimental relationship between the measured RF-dressed splitting and the RF magnetic-field projection amplitude used for the magnetic-field imaging and for the RF-current line-scan analysis described in Sec.~\ref{app:rf_current}.

The RF magnetic-field projection amplitude used for calibration was estimated from the current-induced DC magnetic field using the ratio of the measured RF voltage amplitude,
$V_{\mathrm{RF}}^{\mathrm{pp}}/2$,
to the magnitude of the measured DC voltage.
The DC and RF voltage amplitudes used for calibration were measured at the downstream monitoring point of the sample circuit using the same measurement configuration.
The magnitude of the DC magnetic field projected onto a selected NV axis,
scaled by this measured RF-to-DC voltage ratio,
provided the estimated RF projection amplitude used for calibration.
This procedure uses the measured downstream voltages and does not invoke an independently determined correction for RF impedance mismatch or reflections.

The splitting of the higher-frequency RF-dressed resonance pair was used to calibrate the RF magnetic-field projection amplitude.
Only data points for which the resonance splitting exceeded the mean full width at half maximum (FWHM) of the two RF-dressed resonance dips were included in the linear fit.
For each NV orientation $i$, the splitting was fitted as
\begin{equation}
\Delta\nu_{\mathrm{dress}}^{(i)}
=
\kappa_{\mathrm{exp}}^{(i)}
B_{\mathrm{RF},\parallel}^{(i)}
+
b_{\mathrm{exp}}^{(i)}.
\label{eq:app_rf_calibration}
\end{equation}

Here,
$\kappa_{\mathrm{exp}}^{(i)}$
is the experimentally determined RF field-conversion slope and
$b_{\mathrm{exp}}^{(i)}$
is the fitted intercept.
These quantities are empirical calibration parameters and are distinct from the electron-spin gyromagnetic ratio $\gamma_e$ appearing in the spin Hamiltonian.

The resulting slopes
$\kappa_{\mathrm{exp}}^{(i)}$
were 27.1, 25.2, and 27.3~MHz/mT for NV1, NV2, and NV3, respectively,
and the corresponding intercepts
$b_{\mathrm{exp}}^{(i)}$
were 1.81, 1.50, and 1.72~MHz.
For imaging,
both the experimentally determined calibration slopes and intercepts were used in Eq.~(8).
The same coefficients were also used to convert the RF-dressed splitting measured in the 33.5-MHz RF line scan into the RF magnetic-field projection amplitudes used to determine $I_{\mathrm{RF}}$ in Sec.~\ref{app:rf_current}.

The nonzero intercepts were treated as empirical calibration parameters.
The physical origin of these intercepts and the small differences
among the orientation-dependent calibration slopes remain open
questions beyond the scope of the present study.

\section{Determination of the transverse bias-field component in Fig.~\ref{fig:characterization}(f)}
\label{app:transverse_bias}

The transverse bias-field component
$B_{\mathrm{DC},\perp}^{(1)}$
plotted in Fig.~\ref{fig:characterization}(f)
was determined from the measured NV1 CW-ODMR transition frequencies using the same spin-1 Hamiltonian and strain-averaging procedure described in Sec.~\ref{app:dc_reconstruction}.
This analysis is separate from the determination of the parameters used for the magnetic-field calculations in Fig.~\ref{fig:imaging}.

At each bias-field setting,
the two CW-ODMR transition frequencies associated with NV1 were analyzed.
The transverse strain/electric-field interaction parameters
$E_{x'}$ and $E_{y'}$
were sampled using the same zero-mean Gaussian distributions with a standard deviation of 3.30~MHz described in Sec.~\ref{app:dc_reconstruction}.
For each trial set of parallel and perpendicular bias-field components,
the Hamiltonian was diagonalized and the two predicted transition frequencies were averaged over the strain realizations.

The parallel and perpendicular bias-field components were determined by minimizing the deviations between the strain-averaged calculated transition frequencies and the measured NV1 CW-ODMR resonance frequencies.
The resulting transverse component
$B_{\mathrm{DC},\perp}^{(1)}$
was used for the horizontal axis of Fig.~\ref{fig:characterization}(f).
Thus, the transverse bias field in Fig.~\ref{fig:characterization}(f)
was obtained from the measured CW-ODMR resonances through the spin-1 Hamiltonian rather than from a direct conversion of the mechanical bias-magnet position.

\section*{References}
\bibliographystyle{aipnum4-1}
\bibliography{APL_Eng_Phys_references}

\end{document}